# Deepfake histological images for enhancing digital pathology

Kianoush Falahkheirkhah[1,2], Saumya Tiwari[7], Kevin Yeh[2], Sounak Gupta[8], Loren Herrera-Hernandez[8], Michael R. McCarthy[8], Rafael E. Jimenez[8], John C. Cheville[8], Rohit Bhargava[*,1,2,3,4,5,6]

[1] Department of Chemical and Biomolecular Engineering, University of Illinois at Urbana- Champaign, Urbana, IL 61801
[2] Beckman Institute for Advanced Science and Technology, University of Illinois at Urbana- Champaign, Urbana, IL 61801
[3] Department of Bioengineering, University of Illinois at Urbana- Champaign, Urbana, IL 61801
[4] Department of Electrical and Computer Engineering, University of Illinois at Urbana- Champaign, Urbana, IL 61801
[5] Mechanical Science and Engineering, University of Illinois at Urbana- Champaign, Urbana, IL 61801
[6] Cancer Center at Illinois, University of Illinois at Urbana- Champaign, Urbana, IL 61801
[7] Department of Medicine, University of California San Diego, San Diego, California, 92122
[8] College of Medicine and Science, Mayo Clinic, Rochester, Minnesota, 55905
*rxb@illinois.edu

**Abstract**

A pathologist's optical microscopic examination of thinly cut stained tissue on glass slides prepared from a formalin-fixed paraffin-embedded (FFPE) tissue blocks is the gold standard for tissue diagnostics. In addition, the diagnostic abilities and expertise of any pathologist is dependent on their direct experience with common as well as rarer variant morphologies.  Recently, deep learning approaches have been used to successfully show a high level of accuracy for such tasks. However, obtaining expert-level annotated images is an expensive and time-consuming task and artificially synthesized histological images can prove greatly beneficial. Here, we present an approach to not only generate histological images that reproduce the diagnostic morphologic features of common disease but also provide a user ability to generate new and rare morphologies. Our approach involves developing a generative adversarial network (GAN) model that synthesizes pathology images constrained by class labels.  We investigated the ability of this framework in synthesizing realistic prostate and colon tissue images and assessed the utility of these images in augmenting diagnostic ability of machine learning methods as well as their usability by a panel of experienced anatomic pathologists. Synthetic data generated by our framework performed similar to real data in training a deep learning model for diagnosis. Pathologists were not able to distinguish between real and synthetic images and showed a similar level of inter-observer agreement for prostate cancer grading. We extended the approach to significantly more complex images from colon biopsies and showed that the complex microenvironment in such tissues can also be reproduced. Finally, we present the ability for a user to generate deepfake histological images via a simple markup of sematic labels. These advances allow for building comprehensive diagnostic pipelines that can also handle complex or rare cases, a systematic exploration into tissue complexity by designer histology and a novel tool for pathologist and deep learning.

1. Introduction

The gold standard for histopathologic recognition of most diseases, including nearly all cancers, is the microscopic evaluation of stained tissue sections by a pathologist. This process is now being augmented by innovations in machine learning (ML) for digital pathology, including by deep learning[1–3], to recognize structural and organizational features characteristic of disease. While ML-based workflows can be exceptionally powerful for digital pathology, their development is challenging and uses are limited due to two factors. First, a very large data set needs to be carefully curated so that sufficient examples of rarer diseases or unusual mimics of common disease are included. Consequently, while ML accuracies for common disease conditions (e.g. cancers detection on a whole slide image) is often high, performance for unusual or rare cases may not often be evaluated, be deficient in accuracy, or be inconclusive. Second, ground truth annotations must be supervised by expert pathologists. Annotations for training ML methods are typically more complex than routine clinical diagnosis, expertise may not be readily available, and the extra effort adds to the already strained pathologist workloads. Moreover, such annotations may not be universally agreed upon and require consensus opinions or other statistical approaches to address inter- and intra-pathologist variation. The absence of large, well-annotated data sets that include a sufficient diversity of conditions limits our ability to provide robust ML pipelines for digital pathology. Recent efforts have focused on generating varied synthetic histopathologic images using generative modeling[4,5]. In general, these methods rely on knowledge of glandular structure and predictions based on known patterns (i.e., from the latent space of a deep learning algorithm) that typically focus only on epithelial and stromal classes in a two-phase model of tissue. This approach implies that the generated images can be realistic and large[6]; however, current methods cannot generate intermediate and difficult diagnostic cases, neglect the composition of the microenvironment, and cannot account for subtle changes in the microenvironment associated with the development of disease. The current approach with inclusion of such capability would likely require exponentially more training of the model and data, although initial attempts have been made used parametrized measures[7]. Finally, current approaches do not yet provide the ability to generate "at will" morphology, i.e. allow a directed change to images in morphology. Here we report the generation of realistic histopathologic images that allow precise control over the architecture of tissue, generation of microenvironments with an abundance of cell types, and a facile means to engineer structural and organizational features characteristic of common or rare disease states.

We propose to develop deepfake histological images. Recently, deepfake technologies have provided an array of realistic images, videos, media, and advertisements[8–10] and shown potential for photo editing, entertainment, and, education[11]. Among the most powerful deepfake techniques for synthesizing highly realistic images that can be indistinguishable from actual samples are

Generative Adversarial Networks (GAN)[12], a subclass of deep-learning models. GAN-based models consist of two convolutional neural network (CNN) models, with the first responsible for generating images and second for discriminative feedback to critique the generator via a loss function. In tandem, they work to progressively enhance the realism of the generated images. GAN-based methods have been effectively used to synthesize images for improving medical imaging analysis. For example, in classification tasks as a data augmentation techniques in MRI[13], CT[14], dermatologic conditions[15], and age-related macular degeneration[16]. In histopathology, similar approaches have been investigated to improve the accuracy of classifiers and intelligent augmentation of data in a variety of tasks such as nuclear segmentation[17], identifying cervical intraepithelial neoplasia[18], breast tumor detection[19], and ovarian carcinoma classification[20]. While high quality synthesized images could be obtained, a crucial gap remains in the ability to synthetically control the abundance and characteristics of specific histological units (e.g. epithelial cells or stromal cells). No studies have been reported for the generation of specific pathologies. Furthermore, most synthetic histological images are typically limited (~250 μm x 250 μm) and another gap lies in the ability to synthesize histological images at a larger field of view where multi-cellular morphology or the microenvironment can be appreciably linked to disease or developmental processes. To address these needs, we propose that adding histopathologic identity, i.e. "class labels" can allow for a precise and robust method to tailor synthetic pathology images. Inspired by state-of-the-art research for image editing and manipulation[21], we developed a generative adversarial network (GAN) that accepts class labels as an input and synthesizes histological images (Supplementary Fig. 1), hence termed SHI-GAN. We demonstrate the abilities of SHI-GAN in synthesizing realistic and high-resolution images of colon and prostate tissues, which are two of the most common cancers. Finally, our synthesized images are evaluated by a group of experienced pathologists to assess the quality, accuracy, and reliability of the SHI-GAN framework.

## 2. Methods

### 2.1. Data preparation

This network was trained using two different data sets of Hematoxylin and Eosin (H&E) stain images from prostate and colon samples. The images of prostate samples were downloaded from a publicly available dataset[22]. The dataset included 102 slides of prostate cancer that were scanned using a 3DHistech Pannoramic Flash II 250 scanner at 20x magnification[22]. The dataset included two class annotations epithelial and non-epithelial cells, but in order to localize the nuclei in synthetic images, we added a nuclei class to the dataset by applying a color deconvolution filter[23] with thresholding to separate the resulting Hematoxylin channel and added it to the semantic labels map. A median filter with a 3-pixel kernel was applied to remove salt and pepper noise. For training the framework, 20 biopsies were chosen, split into 25,044 patches of 1,024 x 1,024 pixels each. A second dataset was used to further test the developed approach, which comprised of 1 mm diameter samples across eight tissue microarrays (TMA)[24] that has been previously examined by both FT-IR and optical imaging.[25] Histological segmentation information on this dataset comprised ten histological classes, labeled as epithelium (mature), mucin, epithelium (proliferative), necrosis, reactive stroma, blood, inflammatory cells, non-reactive stroma, muscle and loose stroma with serial sections used to obtain H&E and IR data. The H&E slides were imaged by a Hamamatsu Nanozoomer equipped with a 0.7 numeric aperture (NA) microscopy objective resulting in a pixel resolution of 1 μm. Infrared (IR) images were collected on Agilent Stingray imaging system in high magnification mode with 1.1 μm pixel size at 4 cm$^{-1}$ resolution.

In each case, IR images were annotated by evaluating H&E images after consulting experienced pathologists. We used the output of classifier as the input for the GAN framework which in turn synthesizes the stained images accordingly, which are localized by histological classes. Patches of 512 x 512 were extracted and augmented by random rotation and reflection to create a dataset with 1,632 image patches for training and 500 image patches for testing.

## 2.2. Network's architecture

We implemented a GAN framework largely based on an architecture[21] that includes conditional normalization layers to convert semantic segmentation masks to real images. The architecture of the generator is illustrated in Supplementary Figs. 2a, where a latent space containing 256 components sampled from N (0, 1) is sent to a linear neural network and expanded to 16,384 components. Then, these components are reshaped to 1,024 feature maps of size 4 x 4, followed by several residual blocks (Supplementary Figs. 2b), and finally concluded with nearest neighbor up-sampling layers. At the end of the generator, there is a convolution layer followed by a hyperbolic tangent activation function. The residual blocks include conditional normalization layers[21] (NL) that are described in Supplementary Figs. 2c. To further stabilize the training, we apply a spectral normalization[26] to all the convolutional layers in the generator and discriminator. All of the convolutional layers have kernel size of 3 x 3. Our discriminator architecture largely follows the design of pix2pixHD[27] and can be found in Supplementary Fig. 3, where a series of convolution layers is followed by an instance normalization[28] and LeakyReLU activation function with a parameter of 0.2. This kind of discriminators are computationally less expensive when compared to those with a fully connected network.

## 2.3. GAN framework training details

We trained our generator with the same objective function used in pix2pixHD[27] with a combination of perceptual loss and GAN loss. Perceptual loss is achieved by computing the L1 loss of the feature maps resulting from the last convolutional layer of the pre-trained VGG-19[29] and has been demonstrated to be helpful in image processing tasks such as image super resolution[30]. For calculating the GAN loss, we use 2 discriminators that have identical layouts (Supplementary Fig. 3) applied to different image scales. One operates on full resolution images while the other operates on images down sampled by factor of 2. These multiple discriminators reduce the introduction of unusual and replicated patterns in the synthesized images (Supplementary Fig. 4). The GAN loss is calculated using least square loss term.

We perform 200,000 iterations for each previously described experiment on two NVIDIA 2080 GPUs. Both the generator and discriminator were initialized using Glorot[31]. Adam[32] was used to optimize the parameters of both generator and discriminator with an initial learning rate of 0.0002 while they were multiplied by 0.95 after every 1,000 iterations. Since we trained the networks on three GPUs, we used the synchronized version of the Batch- Normalization. The framework is implemented in PyTorch 1.3, CUDA 10.1, and Python 3.7.1. After training the network and synthesizing a sample set of histological images, the performance was evaluated via survey.

### 2.4. Pathologists review

Expertly trained urologic pathologists reviewed the network synthesized histological images compared to images of real samples in order to evaluate their perceived visual fidelity. Four Pathologists completed the survey. The range of practical experience of our pathologist ranges from 4 to 25 years. We randomly selected 160 real images and corresponding synthesized images to create 8 surveys. Among those real images 80 images have been selected from 2-class model and the rest from 3-class model. Each survey includes 40 images that have been chosen randomly. For each image, the pathologists were asked three questions. The first – "is the image real or synthesized?" – evaluates the ability of the network to synthesize histological images indistinguishable from actual patient samples. The second – "Please rate the image quality (1 is low and 5 is high)" – to establish a metric for evaluating the separability, gradeability, and visual quality of images in either set. In the last question – "What is the Gleason grade?" – we evaluate the synthesized images to assess the potential impact of artifacts on interpretive accuracy.

### 2.5. Statistical analysis

The statistical analysis for the survey is performed separately for each question. Firstly, by inquiring whether the pathologists believe the image is real or synthetic, we evaluate the accuracy, precision, sensitivity, and specificity of the responses. Secondly, by asking the pathologists to provide a quality score for each image, we see if the network output is separable according to this gradation by calculating the average quality score assigned to the real and synthesized images as assessed by each pathologist. Lastly, by obtaining estimates for the Gleason grade, we evaluate the agreement of each pathologist against the reference standard using Cohen's and Fleiss' kappa statistic. While Cohen's kappa statistic is a metric to determine inter-rater reliability between two raters[33], while Fleiss' kappa statistic measures the agreement between three or more raters[34]. We define the consensus reference as the majority diagnosis for each real image. We further calculate Fleiss' kappa statistic, the inter-observer agreement for the real dataset and the synthesized dataset, to compare the concordance rate for the real and synthesized images.

### 2.6. Evaluation using semantic segmentation models

We developed three semantic segmentation models on the prostate cancer dataset in order to evaluate the performance of our framework in synthesizing realistic images. Unlike the GAN framework which takes semantic segmentation as inputs and generates H&E images, these semantic segmentation models take H&E images as inputs and produces the semantic segmentation information. The first model (S1) has been developed using only synthesized images while the second model (S2) used only real data. Finally, a model (S3) has been developed that utilized both real and synthesized dataset. For training the S1 and S2 models, 2,500 image patches of 512 x 512 pixels have randomly been selected for training. Since S3 utilized both real and synthesized datasets, it was trained on 5,000 image patches. For testing and evaluating these models, we used 2,500 image patches from real patients not included in the original training set.

The architecture[35] used for training the semantic segmentation models is a combination of U-Net[36] and ResNet[37] that includes three down-sampling blocks, a bridge block, and three up-sampling blocks. The down-sampling block consists of a convolution layer, a residual block, and a pooling layer at the end. While the down-sampling block reduce the spatial dimensionality, it increases the number of feature maps by factor of two. The bridge block and up-sampling blocks follow the same layout as down-sampling block; however, the bridge block excludes the pooling layer and the up-sampling block, instead substituting it with a bilinear up-sampling layer. The final layer of the architecture includes a convolutional layer to produce 4 feature maps

We used the SoftMax cross entropy objective function to train the semantic segmentation models. The models were initialized randomly and optimized using Adam[32] with initial learning rate of $10^{-4}$. However, the learning rate decreases during training by a factor of 0.95 per 1,000 iterations over a total of 100,000 iterations. A random region of 256 x 256 is chosen at each iteration from 512 x 512 image patches. Training was performed with batch size of 10 on a single NVIDIA 2080 GPU and the framework was implemented in PyTorch 1.3, CUDA 10.1, and Python 3.7.1.

To evaluate the semantic segmentation models, we used standard metrics such as pixel accuracy (PA) and intersection over union (IOU). The pixel accuracy is defined as:

$$PA = \frac{TP + TN}{TP + TN + FP + FN} \tag{1}$$

Where TP, TN, FP, and FN represent the true positive, true negative, false positive, and false negative fractions respectively. However, pixel accuracy is known to be a misleading metric specifically when the classes are imbalanced. An alternative metric to better overcome this challenge is IOU which is defined as:

$$IOU = \frac{TP}{TP + FP + FN} \tag{2}$$

We also report the mean pixel accuracy (mPA) and mean intersection over union (mIOU) which is simply an average of PA and IOU values across all classes.

## 3. Results

As described in the methods section, a complete workflow of network training, image generation and assessment was implemented. To test the capabilities of our approach, we designed scenarios that systematically assess the output of this workflow using increasing capability in reproducing tissue architecture and morphology as a function of the need for complexity and sophistication in input data.

### 3.1. Synthesizing prostate H&E images

In the first assessment, we examined the potential of our approach in using simple annotation to recover complex patterns. Similar to current capabilities of state-of-the-art synthetic histologic image generation, Figure 1a shows the application of our approach to a simple 2-class prostate cancer dataset with 1 μm resolution and patch size of 512 x 512. Since there are two classes, epithelium and "others", nuclei are not separately treated as a class. The advantage of this formulation is that users do not have to address nuclear morphology and positions; the network learns nuclear size distribution and locations. Comparing the synthesized images with ground truth H&E images, we observe a striking similarity in architectural patterns and realistic nuclear morphology. While realistic morphology is obtained in the synthesized images, however, the details of the organization of the nuclei and their specific size distributions are nevertheless derived from an ensemble that approximates the diversity of the training set and cannot precisely predict such structure from the limited input data for individual samples. As an example, the zoomed sections show epithelial cells with more nuclei and do not accurately reproduce the extended cytoplasmic regions, instead providing a greater nuclear density. Thus, these generated images may be useful for mimicking glandular patterns that can be used for training networks for detection and grading of cancer but are ultimately limited for reproducing finer detail. Since a significant diagnostic ability in prostate tissue depends on glandular patterns, these images may be sufficiently useful. Relatively straightforward additions to this basic approach can extend our ability to make predictions more sophisticated. Next, we consider a 3-class prostate tissue dataset with the same 1 μm resolution and patch size of 512 x 512 but now add nuclei as a separate class to assess predictions of their position, size, and shape. Fig. 1b shows the performance of this 3-class model. Both synthesized and real images have the same semantic information and a significantly improved representation of structure is seen, improving both in glandular and nuclear fidelity. While there is expected to be some variation in the predicted images, there are also signs of specific, systematic failures in rare cases. For example, when a group of nuclei are clubbed together (indicated by arrows in zoomed in sections of Fig. 1b) a pattern likely arises in these data that was not effectively reproduced. Though these rare occurrences are known to practicing pathologists, they may or may not arise in a specific training data set. This illustrates a major challenge in synthesizing a diversity of images computationally. These examples do not pose any limitations to the methodology but do point to the need to augment computational models with as many exceptions to the routine such that predictions improve over time.

The results can be improved by both better data and models. One route to better morphological detail is to simply increase the quality of the data. We used the same 3-class prostate tissue annotation but now with 0.5 μm resolution and a patch size of 256 x 256. The results (Fig. 1c) demonstrate greater fidelity in image morphology at all levels and a significantly improved reproduction. Increasing the data quality with emerging, high-resolution and 3D imaging tools can greatly improve the realism and fidelity of this approach. The same framework is capable of generating histological images at different resolutions and fields of view, thus providing a customizable and flexible tool. A second route to higher fidelity images is to examine the input information content. A publicly available prostate tissue dataset[22] was used to develop a semantic segmentation model that classifies H&E images of prostate tissue into epithelial and non-epithelial classes. This work builds on that approach to synthesize additional H&E images from those semantic labels as a comparison to the state of the art. The same approach, however, can be nested to further sub-classify features of the semantic segmentation and build greater complexity as we will discuss later. Before adding to complexity of the model, we describe the results of evaluation of synthesized images.

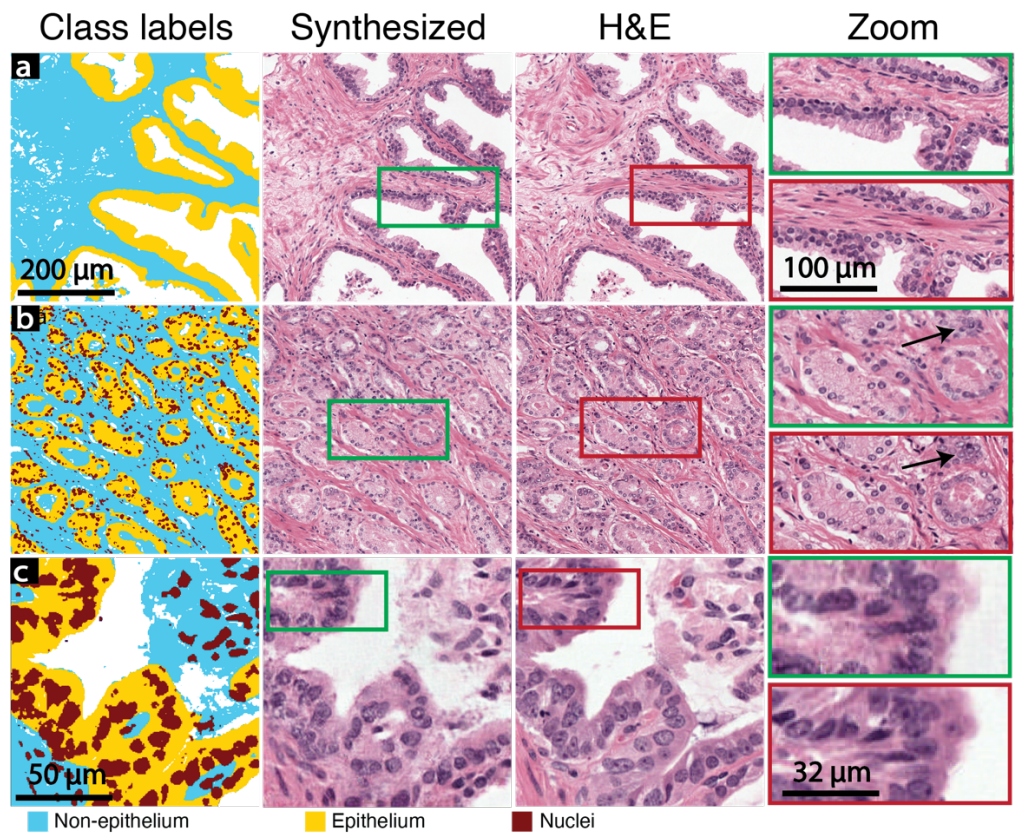

Figure 1. Synthesized and real histologic images of prostate tissue. a, Performance of 2-class model and b, a 3-class model, both with 1.0 μm pixel sizes in the input image. c, Performance of the 3-class model with 0.5 μm pixels. From left to right we show class labels, synthesized images, real images, and zoomed area. A green border denotes the magnified synthesized image, whereas the red border highlights the corresponding real image.

### 3.2. Assessment of generated images

While the capability to synthesize images is powerful, we focused first on assessing whether it may be useful. For each prostate tissue dataset, we developed 3 semantic segmentation predictors using (1) synthetic data, (2) real data, and (3) synthetic+real data in order to evaluate whether a classifier developed using synthetic data is capable of being applied to real data. After training the models, we created a test set that includes 2,500 image patches of 512 x 512 pixels, which did not exist during training the GAN or semantic segmentation models. We used mean pixel accuracy

(mPA) and mean intersection-over-union (mIOU) to evaluate performance across all classes. Table 1a shows the quantitative performance on the test set for 2-class models where mPA and mIOU values for synthetic data are 0.9065 and 0.6902, respectively, which are close to the values achieved for the model developed using real data (0.9104 and 0.6911, respectively). These results suggest that synthetic data generated by our GAN framework performs very similar to real data in training a deep learning model, which implies the synthetic images capture essential elements of tissue structure to a similar degree. Further, developing a segmentation model by merging synthetic and real data yields the highest values for mPA and mIOU (0.9438 and 0.7102, respectively), suggesting augmentation with synthetic data improves performance and corroborates previous studies[13,18]. A similar trend was found for 3-class models, as shown in Table 1b.

| a | 2-class model | | |
|---|---|---|---|
| | Synth | Real | Real + Synth |
| mPA | 0.9065 | 0.9104 | 0.9438 |
| mIOU | 0.6902 | 0.6911 | 0.7102 |

| b | 3-class model | | |
|---|---|---|---|
| | Synth | Real | Real + Synth |
| mPA | 0.9190 | 0.9243 | 0.9522 |
| mIOU | 0.6995 | 0.7018 | 0.7265 |

Table 1. Quantitative evaluation of semantic segmentation models developed using real, synthesized, and real + synthesized. a, 2-class dataset. b, 3-class dataset. The values in the table are mean pixel accuracy (mPA) and mean intersection-over-union (mIOU) that have been averaged across all classes. For both mIOU and mPA, higher is better.

While this first step of assessment used objective, computerized methods to evaluate the quality of synthesized images, we also developed a battery of tests to assess their practical value by domain experts. Four experienced urologic pathologists assessed the utility of this approach via a survey that includes three questions (details in methods). Using 160 images and 160 synthesized images with the same semantic labels, the average accuracy of pathologists in distinguishing real images from synthesized images was found to be ~55% (table 2a). Slightly better than random assignment, this additional ability may come from the multiscale textures both within and around epithelial cells that are not captured by our framework but intuitively appreciable by human cognition. Following this assessment of realism, we focused on assessing the visual quality of images by pathologists assigning a score (1 being low quality to 5 being high quality) to each image. This score focuses on utility of the image, which is related to gradeability, artifacts, and morphology of real and synthesized images. The average visual quality for real and synthesized images was found to be 4.52 and 4.37 respectively (Table 2b), suggesting no major differences between real and synthesized images for histopathologic recognition. Finally, we evaluated the performance of pathologists for assigning a Gleason score. We calculate accuracy and Cohen's kappa ($\kappa$) for pathologists' predictions for synthesized images using a consensus reference standard. The consensus reference standard is determined by the majority of Gleason scores assigned to each real image. Table 2c shows accuracy and $\kappa$ for each pathologist, with median values of 0.899 and (0.716, CI = 95%), respectively. Since $\kappa$ values are a surrogate for agreement that is not due to chance (0 is no agreement, 0.01–0.20 is none to slight, 0.21–0.40 is fair, 0.41–0.60 is moderate, 0.61–0.80 is substantial, and 0.81–1.00 is almost perfect[33]), our results indicate a substantial agreement. In addition to the consensus gold standard, we also evaluated inter-observer agreement for the images using Fleiss' kappa. Fleiss' kappa value for real and synthesized datasets is (0.7219,

CI = 95%) and (0.6594, CI = 95%) respectively, that demonstrates substantial concordance for both datasets.

**a** | **Pathologists performance for recognizing real and synthesized images** | | | |
--- | --- | --- | --- | ---
 | 1 | 2 | 3 | 4
Accuracy | 0.5337 | 0.5405 | 0.6284 | 0.4493
Precision | 0.5474 | 0.5270 | 0.6376 | 0.4734
Sensitivity | 0.4967 | 0.9669 | 0.6291 | 0.7086
Specificity | 0.5724 | 0.0956 | 0.6276 | 0.1793

**b** **Pathologists assessment for comparing visual quality of real and synthesized images (out of 5)** | | | | | |
--- | --- | --- | --- | --- | ---
 | | 1 | 2 | 3 | 4
Real | Mean | 4.4701 | 3.7483 | 4.9470 | 4.9403
 | Std | 0.5866 | 0.5912 | 0.2526 | 0.3313
Synthesized | Mean | 4.2483 | 3.4758 | 4.8069 | 4.9379
 | Std | 0.7123 | 0.7177 | 0.4297 | 0.2692

**c** | **Pathologists performance for assigning Gleason score to synthesized images** | | | |
--- | --- | --- | --- | ---
 | 1 | 2 | 3 | 4
Accuracy | 0.9200 | 0.8780 | 0.8455 | 0.9267
Kappa | 0.7319 | 0.7001 | 0.6878 | 0.7343

Table 2. Pathologists' review as an assessment of synthesized image quality and utility. a, Pathologists' performance in distinguishing real and synthesized images. b, Quality assessment of real and synthesized images (minimum score is 1 and maximum is 5). c, Pathologists' performance for assigning Gleason score to synthesized images. Numbers in the second row of each table indicates pathologists' ID.

### 3.3. Synthesizing H&E Images of Complex Tissues – Colon histopathoplogy

Characterizing the framework using the prostate dataset is both of practical importance but also a good starting step in the broad area of synthesizing H&E images. A significantly more complex morphology, in terms of cellular composition and organization, is found in colon tissue. In colon cancer progression, further, substantial architectural and morphologic changes as well as variations in multiple cell types are observed. To assess the capability of our framework for complex histopathology, we developed a new model to synthesize colon H&E images with 10 important histological classes labeled as epithelium (mature), epithelium (proliferative), necrosis, mucin, muscle, blood, inflammatory cells, non-reactive stroma, reactive stroma, and loose stroma. For colon histology, segmentation data is difficult to obtain from simple H&E images and pathologist annotations are more laborious, while containing the possibility of a higher source of error. Hence, we used IR spectroscopic imaging and its ability to provide highly accurate histopathologic recognition by measuring the intrinsic chemical contrast of tissue[38]. Briefly, IR and H&E images are acquired from serial sections of tissue. Expertly trained pathologists annotate H&E images, which are used as ground truth for assessing results of classification of IR images by machine learning algorithms. Validated algorithms can be used to segment a large number of images. Consequently, we obtained a substantially accurate recognition of cell types, objectively and without detailed pathologists' annotations for every sample. A visual comparison of synthetic image and corresponding H&E images for colon tissue (Fig. 2) shows that the model is capable of generating realistic images in a variety of histological units. For example, Fig. 2a shows muscle tissue surrounded by moderately differentiated adenocarcinoma and reactive stroma whereas Fig. 2b highlights the irregular tumor within reactive desmoplastic stroma. In Fig. 2c, we demonstrate a well differentiated gland which includes mature epithelium (goblet cells) containing cytoplasmic mucin. The success in synthesizing H&E images, given class labels, can lead to creating datasets with varied types of colorectal adenocarcinomas. For instance, mucinous adenocarcinoma is unusual and defined by the presence of more than 50% extracellular mucin associated with neoplastic epithelial cells[39].

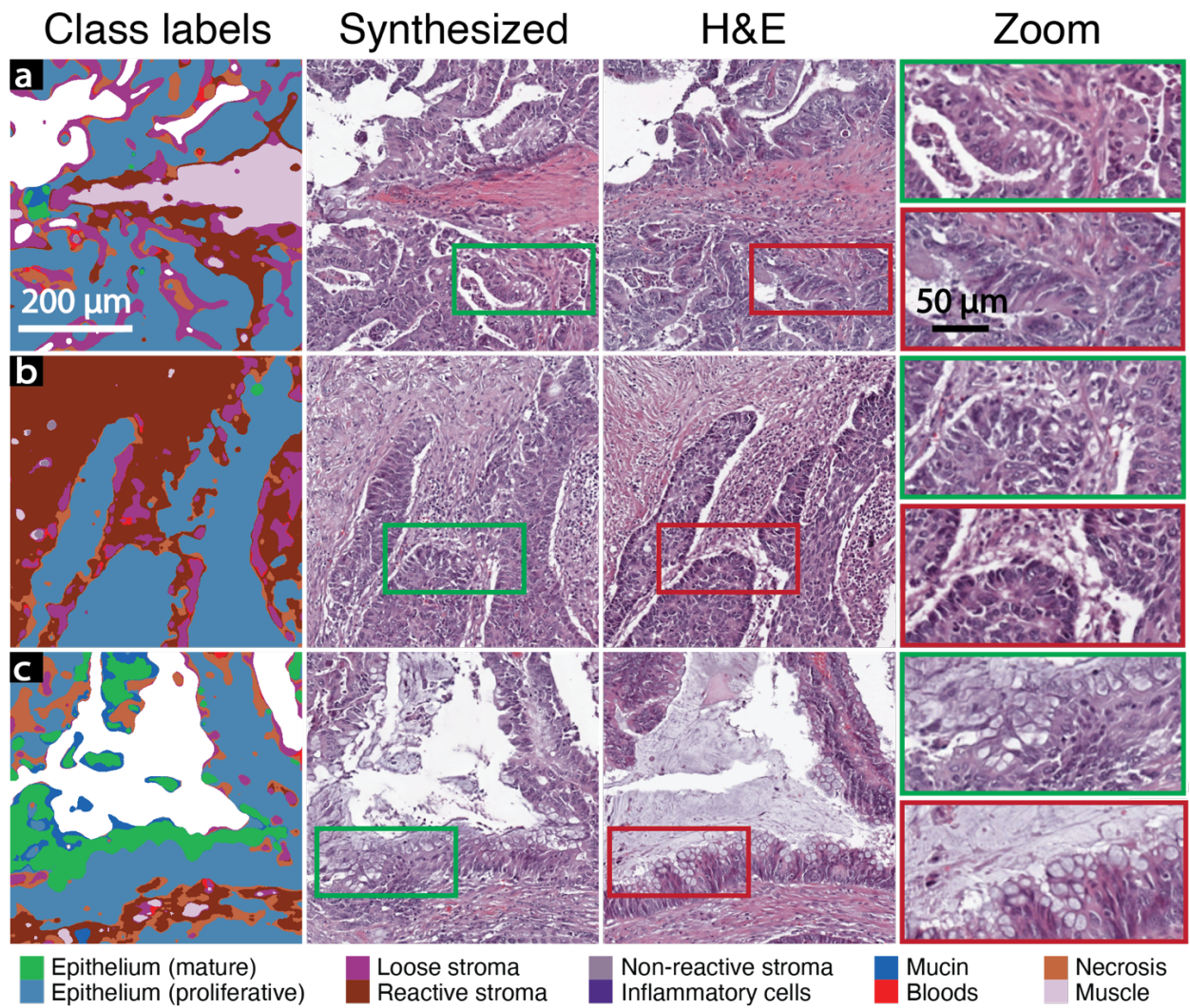

Figure 2. Comparison of synthesized and real images for colon tissue. a, b, and c show different conditions and morphologies, indicating a diverse capability in prediction. Serial sections were used here to obtain H&E images and chemical images with class labels, indicating a small difference between the serial sections.

### 3.4. Generating tissue architecture and organization at will – deepfake histology

While class labels guide our generator in localizing specific histological units, training these models also implicitly learns structural similarities between images. Simplified data representations for the purpose of finding patterns are available in the latent space of our network.

By sampling the latent space repeatedly from a standard gaussian distribution, we generate histological images with different appearances that closely resemble different histological subtypes of cancer. The latent spaces associated with known pathologic conditions can be used to generate images that have features of both, providing a tunability to generating images that may not be naturally abundant. In Fig. 3, we provide an example of linear interpolation and vector operation on the latent space to generate a range of conditions. The resulting synthetic image from moderately and poorly differentiated tumors' latent spaces can be obtained via a simple vector operation. Here, a moderately differentiated tumor with an associated semantic map is transformed to a poorly differentiated one, with a range of intermediate possibilities. While one use of these could be to study human interpretation ability, another could be to synthesize examples of difficult cases in which concordance rate of pathologists as well as performance of machine learning models decreases. For prostate dataset, we demonstrate this interpolation in Supplementary Video 1.

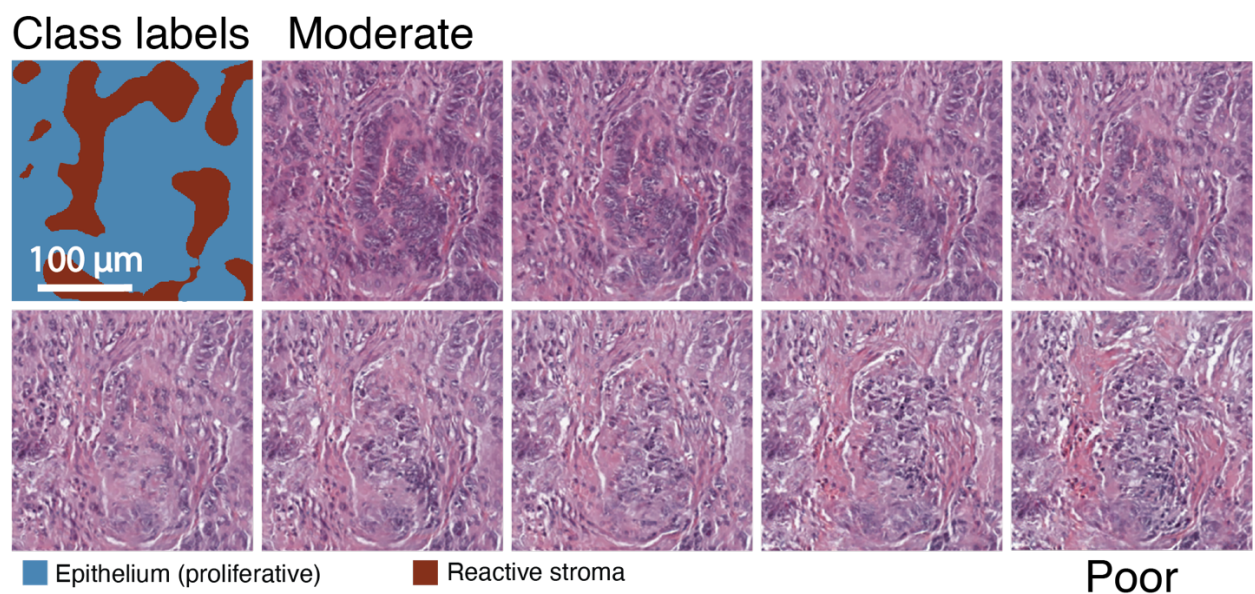

Figure 3. Interpolating latent space to synthesize images between moderately and poorly differentiated colon cancer. All images have the same semantic labels, while they are morphologically different and belong to different grades of cancer.

While simple interpolation between known structures can be powerful, pathologies may not progress along such a continuum and multiple, coupled changes in epithelial and stromal compartments of tissue. At present, there is no clear analytical framework to capture these coupled changes; however, experienced pathologists are able to draw on their expertise or knowledge of specific cases to be able to guide the formation of realistic images. Hence, we extended our framework to design and implement a simple and user-friendly GUI to allow for the merging of human knowledge with abstracted patterns. This GUI allows users to draw arbitrary class labels in a "canvas" mode, to subtly edit existing labels, or to simulate entire transformations such as a desmoplastic reaction and synthesize unique histological examples. For instance, in Supplementary Video 2, we demonstrate the ability of the GUI to draw labels from scratch for 2-class model. As the major metric to determine Gleason grade is the pattern of epithelial regions, the 2-class model is perfect for synthesizing arbitrary pattern of epithelial regions to mimic specific Gleason grade. Similarly, existing class labels can be altered and edited using this GUI as shown in Supplementary Video 3, Supplementary Video 4, Supplementary Video 5 for 2-class, 3-class, and 10-class models, respectively.

## 4. Discussion and conclusion

Here, we report a pipeline to synthesize diagnostic quality images using specific semantic information, using SHI-GAN, which could be used for generating an unlimited number and variety of histological images. These synthesized images can be used to augment image datasets to increase the number of images, increase variation within the dataset to better generalize the model, or generate test cases to assess robustness and performance of learning algorithms. Tissue classifiers developed using synthetic data were shown to have comparable accuracy to classifiers developed using real data, that indicates the realism and quality of synthesized images. In addition to image quality measures, a group of expert pathologists assessed the quality for clinical diagnoses and were able to assign Gleason grade to the synthesized images concordant with real images. This implies that our ability to synthesize images may be useful in testing human ability and for the education of pathologists. Finally, pathologists could identify images as synthesized or real with an ability slightly better than choosing by chance, which has been reported previously[20]. Using higher resolution and including more detailed semantic information of the microenvironment are likely directions to synthesizing even more realistic images.

In order to demonstrate the reliability and generalizability of our approach, we used two different types of tissue, prostate and colon, with different number of histological classes. Unlike the prostate dataset, the colon dataset has significantly more complexity and has class labels that are not locally matched with corresponding ground truth. By using feature-wise, objective perceptual and adversarial loss functions, however, our approach directs the network to generate images that have realistic textural representations even if class labels and ground truth are mismatched at the training stage. This robustness is important since acquisition of paired datasets in biomedical imaging is often challenging and time-consuming. Our framework is not limited to prostate and colon, or H&E stains. To apply the framework to other tissue types, as expected, training on the corresponding dataset will be required to synthesize images. Annotating histologic images is time-consuming, and laborious [17,22]; however, creating datasets using SHI-GAN can reduce the time and labor burden and possibly inadvertent error in annotations. Another interesting aspect of this study is that while we are localizing the histological units, we generate datasets with numerous images of the various prostate cancer patterns alone and in combinations. Such datasets can serve to educate pathologists, test proficiency, and better train robust machine learning models. Finally, the approach overcomes concerns related to patient privacy and data sharing[40] as well as other coordination amongst different centers, as the latent space can be developed locally and extrapolated in a global framework. Since the synthesized images have no direct connection with patients, they can be publicly available and usable with minimal privacy or safety concerns.

Although several approaches have been investigated to overcome the challenges in histological images synthesis, the diversity in tissue makes it very difficult to capture details of multiple cells types and local histological composition. In contrast, SHI-GAN generates images based on class labels, which can be used to incorporate a high degree of complexity computationally as well as provide a tool for interactive image synthesis. While SHI-GAN is unique for histopathology, there are several methods that enable incorporating class labels to the deep learning network to account for synthesizing images for different computer vision tasks. For example, semi-parametric methods[41] can synthesize images given semantic maps but produce less convincing results as they use data extracted from the training dataset itself. In another approach, class labels can be directly input to a deep learning network as an image-to-image translation setting[27,42] that also leverages generative adversarial networks[12]. Again, synthesized images lack the ability to capture sufficient spatial details specific to histologic units and realistic texture. Inspired by previous study[21], our framework synthesize realistic images using a spatially-adaptive normalization, which is a conditional normalization layer that based on input class labels, modulates the activations.

Based on the findings in this study, the framework performs well on the pathologies and tissue architectures that exist in the dataset but has limitations in the realism of images. For example, in prostate dataset, all the non-epithelial histological classes, including luminal secretions, necrosis, stroma, endothelium, and blood cells, are clubbed together. Therefore, the network sometimes synthesizes images that are not in agreement with the ground truth (Supplementary Figs. 6). This limitation of annotation does not pose any limitations to the overall framework since adding more class labels can result in synthetic images that better correspond to the ground truth. Synthesized images reported here were limited in their size (typically, smaller than 0.5 mm x 0.5 mm). There is no theoretical restriction in synthesizing higher resolution and larger fields of view using the methods developed here; the results will scale with computational resources and training time. As with any other machine learning framework, SHI-GAN has some possible failure cases. Our framework does not perform well and sometimes fail to synthesize images if most input class labels belong to any one class (Supplementary Figs. 7), since the method does depend on texture. While this type of situation is not natural, possible failure cases may arise with inexperienced use of our synthesis tool.

In conclusion, we have proposed a GAN-based framework that takes the histological semantic labels and convert the information to photo-realistic pathology images. The synthesized images are largely indistinguishable from real histological images for board-certified pathologists and are diagnostically useful. The generated pathology images can enable new opportunities in developing machine learning tools for digital pathology by providing large data sets, make models more robust by providing unusual cases or intermediate cases, help integrate pathologist knowledge into synthetic images, overcome limitations of patient privacy, and aid education. Advancing generative models for synthetic histology can pave the way for fast, reliable, and efficient diagnoses.